\documentstyle[12pt]{article}

\def\eop{\vspace*{\fill}\pagebreak}

\begin{document}

\begin{titlepage}
{\bf July  1996}\hfill       {\bf PUPT-1632}\\
\begin{center}

{\bf CONFINING STRINGS}

\vspace{1.5cm}

{\bf  A.M.~Polyakov}

\vspace{1.0cm}

{\it  Physics Department, Princeton University,\\
Jadwin Hall, Princeton, NJ 08544-1000.\\
E-mail: polyakov@puhep1.princeton.edu}

\vspace{1.9cm}
\end{center}

\abstract{We propose a hypothesis that all gauge theories are equivalent to a
certain non-standard string theory. Different gauge groups are accounted for by
  weights ascribed to the world sheets of different topologies. The hypothesis
is checked in   the case of  the  compact abelian theories, where we show how
condensing monopole -instanton fields are reproduced by the  summation over
surfaces. In the non-abelian case we prove that the loop equations are
satisfied modulo contact terms. The structure of these terms unfortunately
remains undetermined. }
\vfill
\end{titlepage}
 \par  The theory of quark confinement remains  an unsolved mystery from the
theoretical point of view.  The present status of the subject is the following.
 More than  twenty years ago K. Wilson [1] realized that in the lattice gauge
theory, treated in the strong coupling approximation charges  are confined,
because they are  connected by color -electric strings,  formed by the Faraday
flux lines. Yet this approximation is inadequate for the continuous theory and
thus the problem of confinement is not resolved by this  important result. \par
The next step   was taken by the present author in [2, 3], where the compact
$U(1)$ gauge theories have been analyzed. It was shown that the instanton
configurations of the gauge fields (which were identified as monopoles in $2+1$
dimensions and as monopole rings in $3+1$ )indeed lead to confinement. In the
first case this regime persists for all values of the gauge coupling, while in
the second  it applies only to the couplings above some critical value, due to
the condensation of the monopole world lines. \par Similar conclusions have
been independently reached by t'Hooft [4] and Mandelstam [5] on a qualitative
level. These authors argued that in the "dual superconductor" the vortex lines
must confine electric charges. Of course this picture is identical to the
instanton condensation described above. \par  The quantitative theory developed
for the $U(1)$ case cannot be  generalized  to the non-abelian gauge groups.
This was the rationale to attack the problem from a different angle [6, 7],
where it was conjectured  that there exists an exact duality between gauge
fields and strings. More precisely, it was postulated that there exists a
string theory, governing the behavior of the Faraday flux  lines and thus
equivalent to the Yang - Mills theory.  The analyses was based on  the
equations in the loop space . The intention was to look for their  solution in
the form of string functional integrals. Closely related  ideas were discussed
in [8, 9 ]. \par Although this approach has been developed extensively,  (see
e.g. [ 10 - 13]) the subject is still hovering in limbo. \par To summarize, we
have two approaches to  confinement. One is field -theoretic, trying to
identify relevant field configurations, while the other is based on the string
representation. The main purpose of the present paper is to build a bridge
between these two approaches. \par We will establish the string ansatz in the
case of the abelian gauge groups and make conjectures concerning its
non-abelian generalization. \par Let us begin with a  review of the main points
of ref. [2]. The action in this theory is simple
\begin{eqnarray}
S(A) ={1 \over 4e^{2}}\int{(\partial_{\mu}A_{\nu}
-\partial_{\nu}A_{\mu})^{2}}d\vec{x}
\end{eqnarray}
and the Wilson loop is given by:

\begin{eqnarray}
W(C)=\int{DA_{\mu}{e}^{-S(A)}}\exp{i\oint{A_{\mu}dx_{\mu}}}
\end{eqnarray}
The key point of [2] is that for the compact $U(1)$ one must include the
multivalued monopole configurations of the $A$ fields. As a result the Wilson
loop has a representation
\begin{eqnarray}
W(C)=W_{0}(C) W_{M}(C)
\end{eqnarray}
Here the
 first factor comes from the  Gaussian integration over the $A$ field while the
second factor is the contribution of the point-like monopoles in 3d or monopole
loops in 4d. Let us analyze the 3d case first. A monopole located at the point
$\vec{x}$ contributes to $W_{M}(C)$ as following,

\begin{eqnarray}
W_{M}^{1}(x, C) \propto \exp{(-{const \over e^{2}})}\exp{i\eta(x,C)}
\end{eqnarray}
where
\begin{eqnarray}
\eta
(x,C)=\oint_{C}{A_{\mu}^{(mon)}(x-y)dy_{\mu}}=\int_{\Sigma_{C}}{{(x-y)_{\mu}
\over | x-y| ^{3}}d^{2}\sigma_{\mu}(y)}
\end{eqnarray}
In (4) the first factor is the value of the monopole action , while $\eta (x,
C)$ is the solid angle formed by  point $x$ and the contour $C$.     $ \Sigma
_{C}$  is an arbitrary surface bounded by the contour $C$. As we change this
surface the phase $\eta$ may jump by $2\pi$ leaving $W(C)$ unchanged. As was
explained in [3] , summation over all possible monopole configurations leads to
the following representation:
\begin{eqnarray}
W_{M}(C) \propto \int{D\varphi (x)\exp{-e^{2}\int{[{1 \over 2}(\partial
\varphi})^{2}+m^{2}(1-\cos{(\varphi +\eta)})]d \vec{x}}}
\end{eqnarray}
with $m^{2}\propto e^{-{const \over e^{2}}}$
\par In order to find a  string representation for $W(C)$ we have to recast
these formulae into a different form. Let us consider the following functional
integral,
\begin{eqnarray}
W(C) \propto \int{DB_{\mu\nu}D\phi e^{-\Gamma}}
\end{eqnarray}
with the effective action given by
\begin{eqnarray}
\Gamma =\int{d\vec{x} [{1 \over 4e^{2}}B_{\mu \nu}^{2}+i \phi \wedge dB
+e^{2}m^{2}(1-\cos{\phi})]}+i\int_{\Sigma_{C}}{B_{\mu\nu}d\sigma_{\mu\nu}}
\end{eqnarray}
where we have introduced  a rank two antisymmetric tensor field $B$   . This
field is easy to eliminate by the Gaussian integration.  Let us show that this
procedure returns us to the old expression (6) with an extra factor equal to
$W_{0}(C)$. Thus, this formula unifies the  monopole and the  free field
configurations. Indeed, by minimizing  $\Gamma$ with respect to $B$ we obtain
\begin{eqnarray}
B_{\mu\nu}=i\epsilon_{\mu\nu\lambda}(\partial_{\lambda}\phi
+\int_{\Sigma_{C}}{\delta (x-y)d^{2}\sigma_{\lambda}(y))}
\end{eqnarray}
Using the identity
\begin{eqnarray}
\int_{\Sigma_{C}}{\delta (x-y)d^{2}\sigma_{\lambda}(y)} =\partial_{\lambda}\eta
+\epsilon_{\lambda\rho\sigma}\partial_{\rho}a_{\sigma}
\end{eqnarray}
 where $\eta (x,C)$ is given by (5) and $a_{\mu}(x,C)$ is determined from the
equation
\begin{eqnarray}
\partial^{2}a_{\mu}=\oint_{C}{\delta (x-y) dy_{\mu}}
\end{eqnarray}
we  reduce the integral (8) to the product $ W(C)=W_{0}(C) W_{M}(C) $. We also
notice that $\phi =\varphi +\eta$. Since $\phi$ has no kinetic energy in (8),
it can be eliminated by the Legendre transform, leaving us with the effective
action for the massive $B$ field. As a result we obtain\footnote{Massless
Kalb-Ramond fields, interacting with strings were considered in a different
context in [15]}
\begin{eqnarray}
W(C) =\int{DB e^{-S(B)}\exp{i\int_{\Sigma_{C}}{Bd\sigma}}}
\end{eqnarray}
with the axion action given by
\begin{eqnarray}
S(B) =\int{d\vec{x}({1 \over 4e^{2}}(B_{\mu\nu}^{2}+f(H))}
\end{eqnarray}
Here $H=dB$ and :
\begin{eqnarray}
f(H) =H\arcsin{{H \over m^{2}}}-\sqrt{m^{4}-H^{2}}
\end{eqnarray}
 The multivaluedness of this curious action is a reflection of  $\phi$
-periodicity in  (8). \par Now we confront  a very important puzzle. Namely, we
have  claimed before that the surface $\Sigma_{C}$ is arbitrary.
At the same time if we attempt to compute the Wilson loop by expanding  the
action (14) in $H$ , which is justified for the large enough loops , we find a
non -trivial dependence on this surface. Let us demonstrate this explicitly.
\par In the weak field limit the above action takes the form:
\begin{eqnarray}
S(B)={1 \over 4e^{2}}\int{(B^{2}+m^{-2}(dB)^{2})d\vec{x}}
\end{eqnarray}
In this approximation we obtain:
\begin{eqnarray}
W(C)=e^{-F(C, \Sigma_{C})} \nonumber \\ F =\int_{\Sigma}{d\sigma_{\mu
\nu}(x)d\sigma_{\lambda \rho}(y) D_{\mu \nu , \lambda \rho}(x-y)}
\end{eqnarray}
In this formula the $D$ function is the propagator for the massive axion field
$B$. In the large loop limit we have the following local expansion for  $F$
\begin{eqnarray}
F = c_{1} e^{2}m\int{d^{2}\xi \sqrt{g}} +c_{2}e^{2}m^{-1}\int{d^{2}\xi  (
\nabla t_{\mu \nu} )^{2} \sqrt{g}}  + \ldots
\end{eqnarray}
 where
\begin{eqnarray}
g_{ab}=\partial_{a}\vec{x}\partial_{b}\vec{x}  \nonumber \\ t_{\mu\nu}=\epsilon
^{ab}\partial_{a} x_{\mu} \partial_{b} x_{\nu}
\end{eqnarray}
It  is easy to check that higher order corrections can change the values of the
constants $ c_{1,2}$ but not the structure of this expression. \par This is the
action of the rigid string. What is surprising here is that we have found an
explicit dependence on the surface $\Sigma$ , while originally it was
introduced as an unphysical object. \par The origin of the paradox is the
multivaluedness of the action (14). When expanding it in $H$ we took into
account only one branch of the $(\arcsin{{H \over m^{2}}})$ at each space time
point. The surface independence would be restored had we summed  over all
possible branches. \par The summation over branches can be replaced by the
summation over surfaces. This is the heart of the connection between fields and
strings in this problem. \par Let us begin with proving this fact in the saddle
point approximation. Consider once again the action (8). We have to prove that
there are two equivalent options. First we may allow $\phi$ to vary from
$-\infty$ to $+\infty$ and  minimize this action with respect to $\phi$ and
$B$. The result will depend on $\Sigma_{C}$ but it is easy to see that for two
different surfaces  we have
\begin{eqnarray}
\Gamma (\tilde{\Sigma}) =\Gamma(\Sigma) + 2\pi iN(\Sigma | \tilde{\Sigma})
\end{eqnarray}
with $N$ being an integer. This  follows from the fact that the translation $$
\phi \rightarrow \phi +2\pi $$ insures that $$ dB=0 (\mbox{mod} Z) $$. This is
just another reflection of the fact that  $H=dB$, which measures the departure
from the abelian Bianchi identity, is created by the integer charged monopoles.
\par So  in this first option we get an expression for  $W(C)$ which does not
depend on the shape of the surface but involves all  possible branches of the
$(\arcsin{({H \over m^{2}})})$  summed over separately at each space -time
point -- a rather awkward prescription. \par We claim that an alternative  to
this procedure  is to restrict ourselves to the main branch defined by $$ |
\phi (x) | \leq \pi $$ . After that  the action, which now explicitely depends
on $\Sigma$, must be minimized with respect to this surface. Of course,  as we
go beyond the saddle- point approximation, this procedure must be replaced by
the summation
over surfaces. \par To prove the above statement, let us notice that the
surface variation of $\Gamma$ is given by
\begin{eqnarray}
\delta \Gamma = i \delta \int{B_{\mu \nu}(x)d\sigma_{\mu \nu}(x)}=
i \int{H(x(\xi)) \epsilon _{\mu\nu\lambda}\delta x_{\mu}(\xi)d\sigma_{\nu
\lambda}(\xi)}
\end{eqnarray}
(where we parametrised the surface by  $x_{\mu}=x_{\mu}(\xi _{1},\xi _{2})$ .)
Hence the stationarity condition with respect to the surface is just
\begin{eqnarray}
H(x(\xi))=0
\end{eqnarray}
For a generic function $H(x)$ this equation indeed defines a two-dimensional
surface.  It is convenient now to switch to the  $\phi$
variable. We have

\begin{eqnarray}
-\partial^{2}\phi +m^{2}\sin{\phi} =0 \nonumber \\ H(x)\propto \sin{\phi(x)}
\end{eqnarray}
with the conditions:
\begin{eqnarray}
\lim_{x\rightarrow \infty}{\phi (x)}=0 \nonumber \\ \phi
_{+}-\phi_{-}|_{\Sigma}=2\pi
\end{eqnarray}
where $ \phi _{\pm}$ are the  values of $\phi$ on the two sides of the surface.
This  can also be formulated as a isomonodromy problem for the sine-gordon
equation- we  require  $\phi$ to change by
$2\pi$ as we go around the contour $C$. \par  On the extremal surface,  $\phi$
, which starts from zero, approaches the value $\pi$ on one side of the surface
and jumps to $(-\pi)$ on the other side. This means that on   both sides
$H\propto \sin{\phi}=0$  (which is our stationarity condition). \par If we take
any other surface, the value of $|\phi|$ is necessarily less than $\pi$  on one
side of the surface and greater than $\pi $ on the other side. Hence  in this
case we are pushed out of the fundamental region $|\phi| \leq \pi$. \par To
summarize,  when solving the isomonodromy problem,  the surface of the
discontinuity may be chosen arbitrarily. However among these surfaces the  one
that  minimizes  the action is "more equal than the others" because in this
case the solution lies inside the fundamental region. \par The implication of
this result is very interesting.  It means that at least quasiclassically the
surface  $\Sigma_{C}$  due to its connection with the  $\phi$ -field becomes
dynamical. As a result, some of the degrees of freedom of the $\phi$ field
become stringy. Roughly speaking the dynamical surface appears  as a solution
of  eq. (21). As we integrate over the  $B$ field, we also integrate over
two-dimensional surfaces defined by this equation.
\par  To see this more explicitly, let us return to  eq. (22). Suppose that we
have a solution of this equation. Consider now the spectrum of small
perturbations. It is defined by the following eigenvalue problem,
\begin{eqnarray}
-\partial^{2}\psi _{n}+m^{2}(\cos{\phi _{cl}})\psi_{n}=E_{n}\psi_{n}
\end{eqnarray}
We claim that there are two  classes of the eigenvalues in the case of  the
large loops. The first class includes the ordinary excitations of the field
$\phi$. For them $E_{n}\sim m^{2}$ and they will not be of intrest to us. The
second class is formed by the string modes which have  $$ E_{n} \sim {1 \over
R^{2}} $$ (where  $R$ is the size of the contour.)
\par  The string action (17) must be viewed as a low energy lagrangian for
these modes. In order to check these statements let us recall the solution of
(22) found in (20)  for  the large loops. It has the form,
\begin{eqnarray}
\phi _{cl}=4 sgn(z) \arctan{e^{-m|z|}}
\end{eqnarray}
and  eq.(24) becomes
\begin{eqnarray}
\partial^{2}_{z}\psi+2m^{2}(\cosh{mz})^{-2}\psi =-(E -m^{2})\psi
\end{eqnarray}
(here $z$ -is a coordinate ,normal to the surface.)
It is easy to see that there is a zero mode with $E=0$ given by: $$\psi _{0}
\propto (\cosh{mz})^{-1}$$. Its presence is slightly surprising since the
double layer discontinuity (23) apparently breaks the  translational invariance
and the usual Goldstone -like argument is not directly applicable. However, the
 periodicity of the cosine and the  quantization of the double layer restore
the translational symmetry. \par The above solution is valid for infinite
loops. For finite loops we get  $E \sim R^{-2}\ll m^{2}$ . An approximate
expression for the field $\phi$ accounting for this mode is given by
\begin{eqnarray}
\phi =\phi_{cl}+a(x,y)\psi_{0}(z)
\end{eqnarray}
(where $(x, y)$ are the  longitudinal coordinates on the surface.).
\par  The amplitude $a$ has a natural interpretation as the transverse mode of
string oscillations. The effective action for this mode may be obtained by
passing to the Monge gauge in the covariant action (17).
\par We do not have a  complete proof of this fact. Because of the difficulties
related to the definition of the summation over surfaces (described below in
more details) we will limit ourselves to a somewhat heuristic argument based on
the lattice theory. Let us consider first the partition function, $Z$,  given
by:
\begin{eqnarray}
Z =\int{DB e^{-S(B)}\sum_{(\mbox{all closed surfaces})}{\exp{i
\oint{B_{\mu\nu}d\sigma_{\mu\nu}}}}}
\end{eqnarray}
The latter sum on a lattice can be rewritten as:
\begin{eqnarray}
\sum_{(\mbox{all closed surfaces})}e^{i
\oint{Bd\sigma}}=\sum_{[N_{\vec{x}}]}{\exp{i\sum_{\vec{x}}{N_{\vec{x}}H_{\vec{x}}}}}=\sum_{[q_{\vec{x}}]}{\delta (H_{\vec{x}}-2\pi q_{\vec{x}})}
\end{eqnarray}
Here the  $N$-s and the  $q$-s are  integers.
\par In  (29) we assumed that the surface (which is not necessarily connected)
is represented by a collection of cells in such a way that a cell    centered
at  point $\vec{x}$ is covered $ N_{\vec{x}}$ times.  With this definition of
the surface, it  has no folds, since oppositely oriented components of the fold
cancel each other. At the same time, for simple surfaces without folds and
self-intersections, this definition coincides with the standard one. For
example, a cube located at point $\vec{x}$ corresponds to $N_{\vec{x}}=1$ .
  Eq.(29) is a  consequence of the Stokes theorem.   The monopole
representation of the partition function (6) follows immediately, with the
$q$- s being the monopole charges.  Open surfaces, describing the Wilson loops
are defined and treated analogously. \par This argument is not a proof because
the continuous formulation of the above "fold-less" surfaces is absent. The
main difficulty with such a formulation  lies in the fact that our surfaces
have  an unusual extended reparametrisation invariance which must be
gauge-fixed and renormalized. To see the problem, consider the Wilson loop,
$W=W[\vec{c}(s)]$, for a contour parametrized by $\vec{c}=\vec{c}(s)$. It is
invariant under the  transformations: $$\vec{c}(s)\Rightarrow
\vec{c}(\alpha(s))$$. In the standard string theory we  require that $\alpha $
must be a diffeomorphism, i.e. ${d \over ds}{\alpha}>0$. However, the Wilson
loop is invariant under  the extended reparametrizations for which the
preceding condition is not necessarily satisfied. Hence, for the  confining
strings described by the world surface $\vec{x }=\vec{x}(\xi_{1}\xi_{2})$
we need the action and the measure on the phase space which are invariant under
the extended transformations of the $\xi$ -space, with the Jacobian not
necessarily positive. The string action (12) clearly has this symmetry, but the
measure is still to be defined.
Among other things, this extended symmetry eliminates folds, as we have
already indicated. These problems are not quite solved yet and will be treated
elsewhere. \par In spite of this deficiency, we can draw a number of
interesting conclusions. First of all let us notice that the ultraviolet
cut-off for our string theory is set by the $B$ -field mass, $m$, while the
string tension, $\sigma$,according to (17)  is given by
\begin{eqnarray}
\sigma \propto e^{2}m\sim \alpha M_{W}m\gg m^{2}
\end{eqnarray}
(where we defined a dimensionless coupling constant,  $\alpha={e^{2} \over
M_{W}}\ll 1$ ,and $M_{W}$ is the $W$ - boson mass  coming from the broken
non-abelian theory). The  resonance spectrum is determined by the string
tension, and so we expect to find resonance states in this theory with the
masses $M_{\mbox{res}}$ given by: $$M_{\mbox{res}}\sim (\alpha M_{W}m)^{{1
\over 2}} ; m\sim M_{W}e^{-{\mbox{const} \over \alpha}} $$ $$ m \ll
M_{\mbox{res}}\ll M_{W} $$ \par It is not clear how narrow these states are but
it is obvious that there is an interesting dynamics in the intermediate range
introduced above. \par Second, we conclude that the  confining strings have
mixed nature combining a separate massive $B$ -field with the stringy degrees
of freedom created by the oscillations of the surface $\Sigma$. In the abelian
case treated above they are well separated. \par So far we have  discussed  the
3d abelian theory. It is straightforward to generalize this discussion to the
4d abelian case along the lines of [2]. As was explained there , the confining
phase
related to the condensation of the monopole rings which begins at some critical
coupling, $e_{\mbox{cr}}$, and we have to consider couplings close to this
value. The action (8) is replaced by
\begin{eqnarray}
\Gamma =\int{d\vec{x} [{1 \over 4e^{2}}B_{\mu \nu}^{2}+i \phi \wedge dB
]+e^{2}m^{2}\sum_{\vec{x}\mu}(1-\cos{\phi_{\vec{x}\mu}})}+i\int_{\Sigma_{C}}{B_{\mu\nu}d\sigma_{\mu\nu}}
\end{eqnarray}
Here the field $\phi $ is  a 1-form due to the fact that the instantons in this
case are the monopole rings which are one dimensional objects. We also used the
lattice regularization in this case , which is needed since the coupling isn't
small. All the rest proceeds just as in the previous case. When we restrict the
field $\phi$ to its fundamental domain, $|\phi_{\vec{x}\mu}|<\pi$, we pay for
that with the  summation over surfaces. Again, the field $\phi$ doesn't have
kinetic energy and can be excluded, leading to a certain action for the $B$
-field. Let us notice  that the $\phi$ -field is a dual gauge field interacting
with the monopoles. \par Let us now proceed to the most interesting case of the
non-abelian gauge theories. We would like to put forward a strong conjecture
(already discussed in some form in the previous work[14]). According to this
conjecture all gauge theories are described by the same universal confining
string theory with the action (12). Different gauge groups are accounted for by
 different weights ascribed to surfaces with the  different topology. Thus in
the $U(1)$ -case we have to sum over all topologies indiscriminately, with
equal weights, while in the  $U(\infty)$ -case  only the simplest disk topology
contributes to the Wilson loop. In general,  for the $U(N)$ case we expect the
t'Hooft factor $N^{-\chi}$ ($\chi $ being the Euler character) to appear in the
summation.\par In order to test this conjecture we shall use the loop equations
derived in [7, 10]. The origin of these equations is simple. They arise from an
attempt to find the loop differention which, being applied to the non-abelian
Wilson loop
\begin{eqnarray}
W(C)=\left\langle Tr P\exp{\oint_{C}{A_{\mu}dx_{\mu}}} \right\rangle
\end{eqnarray}
will give the Yang-Mills equations of motion inside the brackets. It is easy to
see that this is achieved by the following differential operation on the loop
space
\begin{eqnarray}
{\partial^{2} \over \partial^{2} x(s)}{}=\lim_{\epsilon\rightarrow
0}{\int_{-\epsilon}^{\epsilon}{dt{\delta^{2} \over \delta x_{\mu}(s+{t \over
2})\delta x_{\mu}(s-{t \over 2})}}}
\end{eqnarray}
The key feature of this operation is
\begin{eqnarray}
{\partial^{2} \over \partial^{2} x(s)}{W(C)}=\left\langle Tr
P(\nabla_{\mu}F_{\mu\nu}(x(s))({d \over
ds}{x_{\nu}})\exp{\oint_{C}{A_{\mu}dx_{\mu}}}) \right\rangle \approx 0
\end{eqnarray}
Here $F_{\mu\nu}$ is the standard Yang-Mills field strength, $\nabla$ is the
covariant derivative, and the sign $\approx$  here and below means  "modulo
contact terms", i.e. terms which are nonzero  only when the correlation
function under consideration contains coincident space -time points.\par One of
the reasons for our universality conjecture is that according to the above
formula the wave operator acting on the Wilson loop is universal, while the
contact terms, closely related to the topology of the surfaces involved, are
sensitive to the gauge group. In what follows we will show that our string
ansatz indeed satisfies eq.(34). In order to prove this statement we have to
derive some variational formulas. Consider a surface parametrised by the
equation
\begin{eqnarray}
\vec{x}=\vec{x}(\xi ; [\vec{c}(s)])
\end{eqnarray}
where we  have  introduced explicit dependence  on the boundary contour which
is parametrized by $\vec{x}=\vec{c}(s)$. Our aim is to apply the wave operator
to our string ansatz. The first variation gives:
\begin{eqnarray}
{\delta \over \delta
c_{\mu}(s)}\int{B_{\lambda\rho}(x)d\sigma_{\lambda\rho}}=B_{\mu\rho}(c(s))\dot{c}_{\rho}(s)+\int{(dB)_{\alpha\beta\gamma}{\delta x_{\alpha} \over \delta  c_{\mu}(s)}d\sigma_{\beta\gamma}}
\end{eqnarray}
The second step is to apply the second derivative ${\delta \over \delta
c_{\mu}(u)}$ and to extract from this expression the term proportional to
$\delta(s-u)$. As is clear from our definition, our loop operator is just the
coefficient in front of this $\delta $-function. Using this fact we obtain:
\begin{eqnarray}
{\partial^{2} \over
\partial^{2}c(s)}\int{B_{\lambda\rho}(x)d\sigma_{\lambda\rho}}={\partial\over
\partial c_{\mu}(s)
}B_{\mu\rho}(c(s))\dot{c}_{\rho}(s)+\int{(dB)_{\alpha\beta\gamma}({\partial
^{2}x_{\alpha}\over \partial^{2}c(s)})d\sigma_{\alpha\beta\gamma}}
\end{eqnarray}
The partial derivative  ${\partial \over \partial c_{\mu}(s)}$ is again defined
as a coefficient in front of the correspnding  $\delta$-function.
It reduces to the ordinary partial derivative when acting on functions, but in
our case it is important to remember  that $B_{\mu\nu}$ is actually a
functional. When deriving these formulae we  have dropped several terms
containing the products of two variational derivatives in the integrand of
(37).  This is allowed because for  smooth surfaces such terms never contain
the  factor $\delta(s-u)$ which we are looking for. Essentially, the formula
(37) means that our wave operator, despite its appearence, is of the first
order and satisfies the Leibnitz rule. The properties of  these loop
derivatives were discussed in ref[7] in more details. \par Let us show now that
if we use the equations of motion for $\vec{x}(\xi)$ and $B_{\mu\nu}(x)$ the
right hand side of (37) is indeed zero modulo contact terms. First of all, as
was already noticed, equations of motion for $\vec{x}(\xi)$ have the form:
\begin{eqnarray}
dB(\vec{x}(\xi))=0
\end{eqnarray}
and thus exterminate the second term in (37). In order to deal with the first
term we need the equations of motion for the $B$- field. If the action has the
form;
\begin{eqnarray}
S(B)={1 \over
4e^{2}}\int{d\vec{x}[B_{\mu\nu}^{2}+m^{-2}(dB)^{2}+\ldots]}+i\int{B_{\mu\nu}d\sigma_{\mu\nu}}
\end{eqnarray}
we get the following classical value of $B$:
\begin{eqnarray}
B_{\mu\nu}(\vec{x} ; \left\{ c(u) \right\}) \propto \int{\delta
(x-y)d\sigma_{\mu\nu}(y)}\nonumber \\ {\partial \over \partial
x_{\mu}}{B_{\mu\nu}(x,c)}\propto \oint_{C}{\delta (x-y)dy_{\nu}}
\end{eqnarray}
We took the limit   $m\rightarrow \infty$  in these formulae, since, as was
explained above, the mass of the $B$-field serves as an ultraviolet cut-off for
the string modes. When computing the derivative ${\partial \over
\partial{c_{\mu}(s)}}$ we have to add the above value of
$\partial_{\mu}B_{\mu\nu}(x,c)$ at the point $\vec{x}=\vec{c}(s)$ to the result
of the variation with respect to $\vec{c}(s)$. Somewhat lengthy computation
shows that in the large mass limit the leading term vanishes :
\begin{eqnarray}
\lim_{m\rightarrow \infty}{{\partial \over \partial c_{\mu}(s)}{B_{\mu\nu}}}=0
\end{eqnarray}
\par All this  means is  that, on a superficial level, our string ansatz
satisfies the loop equations. However, we are far from claiming that our gauge
field -string hypothesis is proved by the above computations. The main
difficulty lies in our present inability to analyze the  contact terms which as
usual appear when equations of motion are used in the functional integral.
Correct gauge fixing and renormalization are crucial for this task.   At
present the best we can do is to hope that these terms are essentially fixed by
the topology of the surface and by dimensional counting. If this is true they
must be the same in the Yang -Mills theory and in our string theory, thus
proving our hypothesis. \par Despite the incompleteness of these results, I
feel that some progress has been made in establishing relations between fields
and strings.
\par I am grateful to A. Dubin, M. Douglas,  V. Kazakov, I. Klebanov, I. Kostov
and E. Witten
for useful discussions. I would like to thank D. Makogonenko for support and
encouragement.

\par This work was partially supported by the National Science Foundation under
contract PHYS-90-21984.

\eop
\appendix{REFERENCES \par [1] K. Wilson Phys. Rev. D10 (1974) 2445\par [2] A.
Polyakov Phys. Lett. 59B (1975) 82 \par [3] A. Polyakov Nucl. Phys. B120(1977)
429 \par [4]
 G. t'Hooft  in High Energy Physics , Zichichi, Editrice Compositori, Bolognia,
(1976)\par [5]S.Mandelstam Phys. Rep. 23C(1976)245 \par [6] A. Polyakov Phys.
Lett. 82B(1979)247 \par [7] A. Polyakov Nucl. Phys.B164(1980)171\par[8] Y.
Nambu Phys.Lett 80B(1979)372\par [9] J.Gervais A. Neveu Phys. Lett.
80B(1979)255\par[10] Y. Makeenko A. Migdal Nucl. Phys. B188(1981)269\par [11]
V. Kazakov Sov. Phys. JETP, 58 (1983) 1096\par [12] I. Kostov Nucl. Phys. B265
(1986) 223\par [13] D. Gross W. Taylor Nucl. Phys. B400(1993)181\par [14] A.
Polyakov in Proceedings of Les Houches School (1992) Elsevier\par[15] M. Awada
D. Zoller Phys. Lett. B325(1994) 115}

\end{document}